\documentstyle[12pt]{article}
\begin{document}
\vspace{-2cm}

\def\til{\smash{\mathop{\sim}\limits_{\raise3pt\hbox{$\scriptstyle z\to 1$}}}}

\def\setazero{\smash{\mathop{\rightarrow}\limits_{\raise3pt\hbox{$\scriptstyle z\to 0$}}}}

\def\seta1{\smash{\mathop{\rightarrow}\limits_{\raise3pt\hbox{$\scriptstyle
z\to 1$}}}}

\def\undertil{\smash{\mathop{\simeq}\limits_{\raise3pt\hbox{$\scriptstyle
x_F \to 1$}}}}

\def\maior{\smash{\mathop{>}\limits_{\raise4pt\hbox{$\scriptstyle \sim$}}}}

\title{\centerline {\bf Asymmetries in Heavy Meson Production}
\centerline {\bf from Light Quark Fragmentation}}

\author{{\sl J. Dias de Deus} \thanks{\it e-mail: jdd@fisica.ist.utl.pt}  \ {\rm and} \ 
{\sl F. O. Dur\~aes} \thanks{\it e-mail: fduraes@axpfep1.if.usp.br}
 \thanks {(On leave from Instituto de F\'{\i}sica, Universidade de S\~ao Paulo - SP, Brazil)} \\[0.1cm]
{\small Departamento de F\'{\i}sica/CENTRA, IST,}\\ 
{\small  1096 Lisboa Codex, Portugal}}

\maketitle

\medskip

\vskip 2 true cm

\vspace{1cm}

\begin{abstract}
We discuss the possibility of the asymmetry in $D^-/D^+$ production, from
$\pi^-$ beams, being a direct consequence of the properties of the light
quark fragmentation function into heavy mesons. The main features of the
asymmetry, as a function of $x_F$, are easily described. An integrated
sum-rule for the $D^-, D^+$ difference, is presented. Predictions for the
asymmetry, in $B$ meson production, are given.\\
\end{abstract}

\vskip 2 true cm

In the framework of perturbative QCD it is not easy to explain the observed
asymmetry in the production of leading and non leading charmed  mesons in
fixed target experiments with $\pi^-$ beams \cite{frix}. In fact, in the $x_F
\geq 0$ region, an excess of $D^- (d\bar c)$ over $D^+ (\bar d c)$ is
observed, the asymmetry, defined as
\begin{eqnarray}
A (x_F) \equiv {N^- (x_F) - N^+ (x_F) \over N^- (x_F) + N^+ (x_F)} \ , \label{eq:eq1}
\end{eqnarray}
increasing with $x_F$. The effect does not show appreciable dependence on
$p_T$.

In QCD, charmed quarks are generated by parton-antiparton fusion and in
this process the $c$ quarks are produced  with relatively small rapidities
such that their fragmentation or coalescence probability is not likely to
reproduce the observed asymmetry \cite{vogbro}. Some models, containing
recombination mechanism \cite{hwa,igm,herr,lik}, fast $c$-quark strings 
\cite{sjos} and intrinsic charm \cite{brohoy,pisk} can be adjusted to reproduce 
data.

In the present paper we argue that the asymmetry may be essentially due to
the, so far neglected, $d\to D^- (d\bar c)$ fragmentation which, for large
$x_F$, gives the required $D^-$ dominance.

Experimentally, not much is known about the production of $D$ (or $B$)
mesons from light quarks. In $e^+ e^-$ annihilation, most of the heavy
mesons come from heavy quarks. Detection of heavy meson in one hemisphere,
with {\bf absence} of heavy meson in the opposite hemisphere, would be an
indication of fragmentation from light quark.

Theoretically, and taking as an example $D^-$ meson production, if one
looks at $e^+e^-$ at $Q=2m_D$ threshold there are two possibilities:
production from $d$ quark and production from $\bar c$ antiquark (Fig.1).
At the $m_D$ threshold the two fragmentation functions, $D_{D^-/d} \, (z,m_D)$ 
{\rm and} $D_{D^-/\bar c} \, (z,m_D)$ are of the form \cite{georg,jddnsak},
\begin{eqnarray}
D_{D^-/d} \, (z,m_D) \sim \delta (1-z) \ , \label{eq:eq2}
\end{eqnarray}
\begin{eqnarray}
D_{D^-/\bar c} \, (z, m_D) \sim \delta (1-z) \ , \label{eq:eq3}
\end{eqnarray} 
with $z=2P \cdot Q/Q^2$, $P$ and $Q$ being the four-momenta of the $D$
meson and the virtual photon, respectively. The normalizations in (2) and
(3) are naturally different, a larger factor is expected in the $\bar c$
antiquark fragmentation. But, in both cases, the $D^-$ meson takes all the
energy available: $z=1$. At the same threshold energy $m_D$ the unfavoured
fragmentation functions $D_{D^-/c} \, (z,m_D)$ and $D_{D^-/\bar d}  \,(z,m_D)$ 
are identically zero.

By making use of threshold energy fragmentation functions and applying QCD
non-singlet evolution and (possible) spin and ressonance effects \cite{jddnsak},
one arrives, at $Q>m_D$, to the fragmentation functions:
\begin{eqnarray}
D_{D^-/d} \, (z,Q) - D_{D^- /\bar d} \, (z,Q) \label{eq:eq4}
\end{eqnarray}
\begin{eqnarray}
D_{D^- /\bar c} \, (z,Q) - D_{D^- /c} \, (z,Q) \ . \label{eq:eq5}
\end{eqnarray} 

The key idea is that these two non-singlet fragmentation functions, as QCD
is flavour independent, remain similar in shape. The functions (4) and (5)
are peaked at an intermediate value of $z$ \cite{jddnsak}, and can be
represented, for instance, by Peterson's parameterization \cite{petzerw}. The
main result is that the $d\to D^-$ fragmentation will produce, like $\bar c
\to D^-$ fragmentation, fast mesons. Note that a similar behaviour is also
observed in $u\to K^+$ and $\bar s \to K^+$ fragmentation functions \cite{feyn}. 

One should notice that in (4), because of charge conjugation, we have the
difference between leading and non-leading $D$ mesons, with $\pi^-$ beams,
as it appears in the numerator of the asymmetry (1):
\begin{eqnarray}
D_{D^- /d} \, (z,Q) - D_{D^+ /d} \, (z,Q) \ . \label{eq:eq6}
\end{eqnarray} 

At large $x_F$ the $D$ mesons must come from a fast quark and $x_F \simeq
z$. This means that, at least for large $x_F$, (6) must behave similarly to
the Peterson's formula. Within large errors, this is consistent with data,
as we shall see later. 

In order to construct the full fragmentation functions, as the perturbative
QCD is not simple and is too ambiguous, concerning final states, we shall
make use of an old non perturbative model of Krzywicki and Peterson \cite{krzyw}, 
further developed in \cite{jdd}. The mesons in the quark
fragmentation are generated via an integral equation and are produced
ordered in rapidity (see Fig.2), such that meson 1 is, on the average,
faster than meson 2, etc. If $D_1 (z)$ is the fragmentation function for
the first rank meson, $D_2 (z)$ the fragmentation function for the second
rank meson, etc, we have \cite{jdd}
\begin{eqnarray}
D_2 (z) = \int_0^{1-z} D_1 (z') D_1 \left( {z\over 1-z'}\right) {dz' \over
1-z'} \ ,& \label{eq:eq7}
\end{eqnarray} 
and, in general,
\begin{eqnarray}
D_k (z) = \int_0^{1-z} D_1 (z') D_{k-1} \left( {z\over 1-z'}\right) {dz'
\over 1-z'} \ . \label{eq:eq8}
\end{eqnarray}

The functions $D_k (z)$ are normalized to $1$. The full fragmentation
function is
\begin{eqnarray}
D(z) = \sum_k D_k (z) \ . \label{eq:eq9}
\end{eqnarray}

As only the leading meson can be of rank 1, the function $D_1 (z)$ is
nothing but the difference between the leading and non-leading
fragmentation functions (see (6)).

It is easily seen that in the limit $z\to 1$ from (7) and (8) we obtain:
\begin{eqnarray}
D_2 (z) \ \til \ D_1 (0) D_1 (z) (1-z) \ , \label{eq:eq10}
\end{eqnarray} 
and
\begin{eqnarray}
D_k (z) \ \til \ D_1 (0)^{k-1} D_1 (z) (1-z)^{k-1} \ . \label{eq:eq11}
\end{eqnarray}

In this limit, keeping only the most important terms, $D_1$ and $D_2$, and
identifying $z\simeq x_F$ we obtain for the asymmetry (1):
\begin{eqnarray}
A (x_F)_{x_F \to 1} = {D_1 (x_F) + D_2 (x_F) -D_2 (x_F) \over D_1 (x_F) +
D_2 (x_F) + D_2 (x_F)} \simeq 1-2D_1 (0) (1-x_F) \ . \label{eq:eq12}
\end{eqnarray}

Two remarks can be made regarding Eq. (12). The first one is that the
asymmetry increases and approaches 1 as $x_F \to 1$. The second one is that
the approach of the $x_F \to 1$ limit is controlled by the behaviour of
$D_1 (z)$ at $z\to 0$. In order to see the importance of this point let us
assume that in the $z\simeq 0$ region the function $D_1$ behaves as: 
\begin{eqnarray}
D_1 (z) \ \setazero \ z^{\alpha} \ , \label{eq:eq13}
\end{eqnarray}
with $\alpha > -1$. The function $D_2 (z)$, in the $z\to 1$ limit, will
behave as 
\begin{eqnarray}
D_2 (z) \ \seta1 \ D_1 (z) (1-z)^{\alpha +1} \label{eq:eq14}
\end{eqnarray} 
and the asymmetry 
\begin{eqnarray}
A (x_F) \ \undertil \ 1-c \ (1-x_F)^{\alpha +1} \ , \label{eq:eq15}
\end{eqnarray} 
$c$ being some normalization constant. By computing the second
derivative $d^2 A/dx^2_F$ one immediately sees that the $x_F \to 1$ limit
is approached with negative curvature if $\alpha >0$, with positive
curvature if $-1 < \alpha < 0$, and in a straight line manner if $\alpha
=0$. As we shall see later, data favours a behaviour of $D_1 (z)$ in the
$z\to 0$ limit with $\alpha \maior 0$, as expected from QCD non-singlet
evolution \cite{jddnsak} and from Peterson's formula \cite{petzerw}.

In order to be somewhat more precise we shall look more carefully to $D^-$
and $D^+$ production from $\pi^-$ beam, keeping the relevant functions, in
the $x_F \to 1$ limit, $D_1$ and $D_2$. We have the contributions of Fig.3.
While contributions a) and b) involve the function $D_1$ with a charm
quark, the contributions c) and d) involve the function $\tilde D_1$
without charm quark ($d\to d\bar u$ or $\bar u \to \bar u d$). The signals
$-$ and $+$ indicate $D^-$ and $D^+$ meson production, respectively, but
$D_1^+ = D_1^-$, etc. The factors $1/2$ account for isospin (strange quarks
are neglected in comparison with $u$ and $d$ quarks). The function $\tilde
D_2$ is written (see Fig.3),
\begin{eqnarray}
\tilde D_2 (z) = \int_0^{1-z} \tilde D_1 (z') D_1 \left( {z\over
1-z'}\right) {dz'\over 1-z'} \ . \label{eq:eq16}
\end{eqnarray}

We can finally write the fragmentation functions $D^-(z)$ and $D^+ (z)$,
keeping only $D_1$, $D_2$ and $\tilde D_2$ contributions, and identifying
again $x_F \simeq z$, as 
\begin{eqnarray}
D^- (x_F) &=\ D_1 (x_F) + {1\over 2} D_2 (x_F) + {1\over 2} \tilde D_2 (x_F)
+ \ldots & \label{eq:eq17}
\end{eqnarray}
\begin{eqnarray}
D^+ (x_F) &=\ \qquad \quad \ \ \quad {1\over 2} D_2 (x_F) + {1\over 2}
\tilde D_2 (x_F) +\ldots & \label{eq:eq18}
\end{eqnarray} 
and, for the asymmetry, 
\begin{eqnarray}
A(x_F) \ \undertil \ {D_1 (x_F) \over D_1 (x_F) + D_2 (x_F) + \tilde D_2
(x_F) +\ldots } \label{eq:eq19}
\end{eqnarray}

In an attempt to compare (19) with experiment we have used for $D_1 (z)$
and $\tilde D_1 (z)$ Peterson's parameterization, with the values 
$\varepsilon = 0.06$ $({<z>_{d\bar c}} \, \simeq 0.78)$ for $D_1$ and 
$\varepsilon \simeq 3.3$ $({<z>_{\bar ud}} \, \simeq 0.35)$ for $\tilde D_1$, 
and (7) and (16) to compute $D_2$ and $\tilde D_2$. Note that the use of Peterson's
formula is theoretically not justified for $d\to d\bar u$ or $\bar u \to \bar u d$ 
fragmentation, the reason why we used it is because it reasonably fits data and 
parametrizations of $\tilde D_1 \equiv 2 (D_{\pi^+/u} - D_{\pi^-/u})$ 
\cite{jddnsak,feyn,jdd}.

In Fig.4 we compare directly our formula (19) with data on the $\pi^- \to
D^{\pm}$ asymmetry. The agreement is reasonable, better than it should. The
limit $A=1$ is approached with negative curvature due to the fact that
$D_1$ and $\tilde D_1$, experimentally and in agreement with Peterson's
formula, smoothly vanish as $z\to 0$ (see (13)). Including higher rank $D$
meson production, i.e., $D_3 , \tilde D_3 , D_4 ,\tilde D_4$, etc.,
fragmentation functions, (19) goes to zero faster than in the figure, as
$x_F$ moves to zero.

It is clear that the comparison of Eq. (19) with data in Fig.4, except in
the $x_F \to 1$ region, is far from being justified. In general, one
requires the convolution of parton structure functions $f(x)$ with
fragmentation functions $D(z)$, with $x_F = zx$, at least for the valence
$\bar u$ and $d$ quarks. In doing so, one realizes that leading particles,
$D^-$, can be produced even at small $x_F$ (from small $x$ quarks) and, in
a pure fragmentation approach, as considered here, the asymmetry is {\bf
not} expected to approach zero as $x_F \to 0$.

We believe that in our comparison with data in Fig.4 we are making two,
somehow compensating, mistakes. Inclusion of $D_3 , \tilde D_3 , D_4 ,
\tilde D_4$, etc., contributions would have decreased the asymmetry for
small $x_F$. Inclusion of simultaneous fragmentation of $\bar u$ and $d$
would have increased the asymmetry for small $x_F$. As we mentioned above
our results are better than they should have been, but we think that we
understand why is it so.

There is a general result, from our approach which is independent of
convolution calculations involving structure functions. If one selects the
sample of events where $D^-$ or/and $D^+$ are produced, then
\begin{eqnarray}
<N^-> - <N^+>\ = \ 1/2 \label{eq:eq20}
\end{eqnarray} 
where $<N^{\mp}>$ is the $D^{\mp}$ average multiplicity in the $D^-, D^+$
sample. If in (20) one takes normal average multiplicities (over all the
events), the right hand side of (20) becomes $1/2 \, \sigma^D /\sigma_{in.}$, 
where $\sigma^D$ is the cross-section for $D^{\pm}$ production and $\sigma_{in.}$ 
the inelastic cross-section. In principle, it is not difficult to test 
experimentally Eq. (20).

In Fig.5 we show our prediction for the $B^- , B^+$ asymmetry. As the $b$
quark fragmentation $D_1$ is, in this case, closer to a $\delta$-function
($\varepsilon = 0.018, {<z>_{u \bar b}} \, \simeq 0.87$, in Peterson's formula) 
the asymmetry becomes important closer to $x_F \to 1$.

Concerning $D^0 , \bar D^0$ production from $\pi^-$ beams an asymmetry
essentially identical to the $D^- , D^+$ asymmetry is expected. In the case
of $D_s^-, D^+_s$ production, naturally no asymmetry is expected \cite{e791}.

Our treatment of light quark fragmentation functions into heavy mesons can
be easily included in multi-interacting parton models, as the Dual Parton
Model \cite{capel}.

\ \bigskip

{\bf Acknowledgments}
\medskip

J.D.D. thanks the hospitality at the Instituto de F\'{\i}sica, Universidade
de S. Paulo, and usefull discussions with Fernando Navarra. F.D. thanks the
hospitality at the Departamento de F\'{\i}sica and CENTRA, Instituto
Superior T\'ecnico, where most of this work was done, and the help from
everybody there. F.D. acknowledgs a research grant from CAPES, Brasil. This
work was supported by the contract PRAXIS/PCEX/P/FIS/124/96.

\vfill \eject

\vfill \eject

\noindent

{\bf Figure Captions}\\
\begin{itemize}
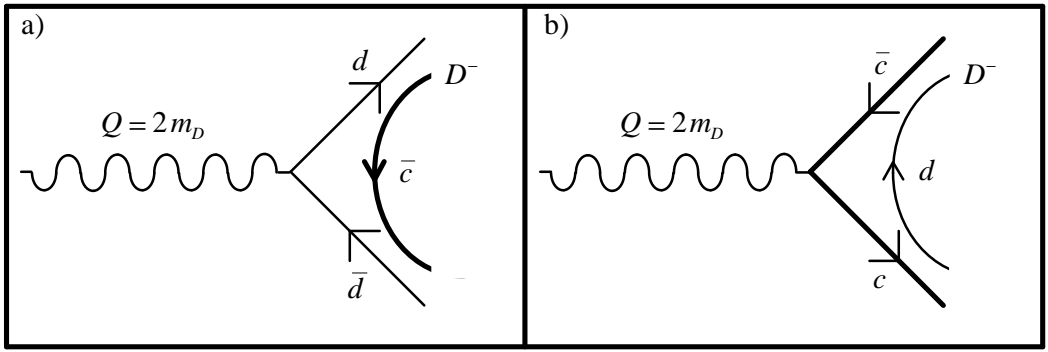
\item[{\bf Fig. 1}] The fragmentation functions, a), $d\to D^-$ and, b),
$\bar c \to D^-$ at the $Q=m_D$ threshold. The two functions have the same
shape in $z$, even after QCD evolution.

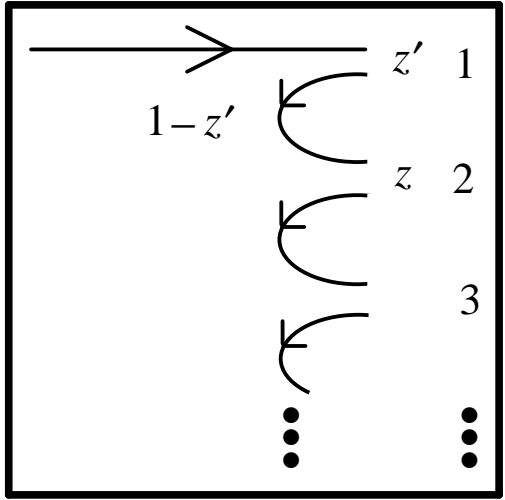
\item[{\bf Fig. 2}] Model to generate fragmentation functions with ordered
mesons, rank 1, 2, 3, etc. See Eqs. (7) and (8).

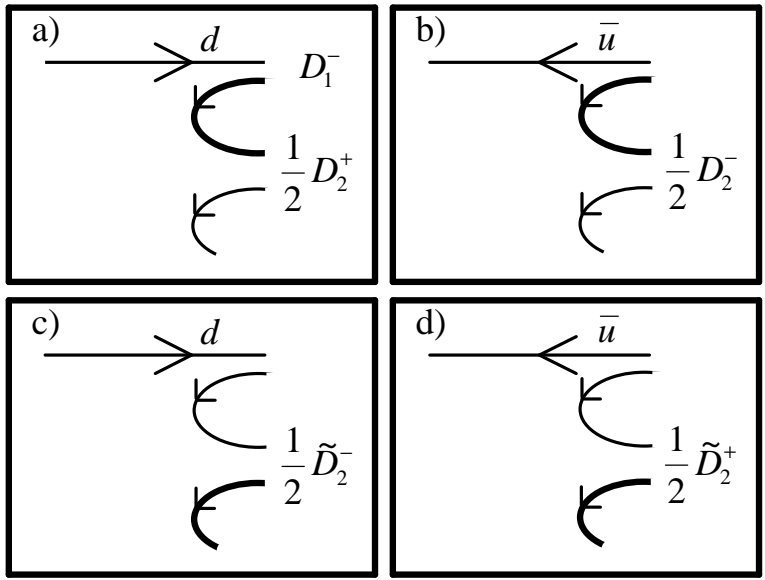
\item[{\bf Fig. 3}] Fast, rank 1 and rank 2, contributions to the $D^{\mp}$
fragmentation functions. Only diagram a) contributes to the asymmetry.

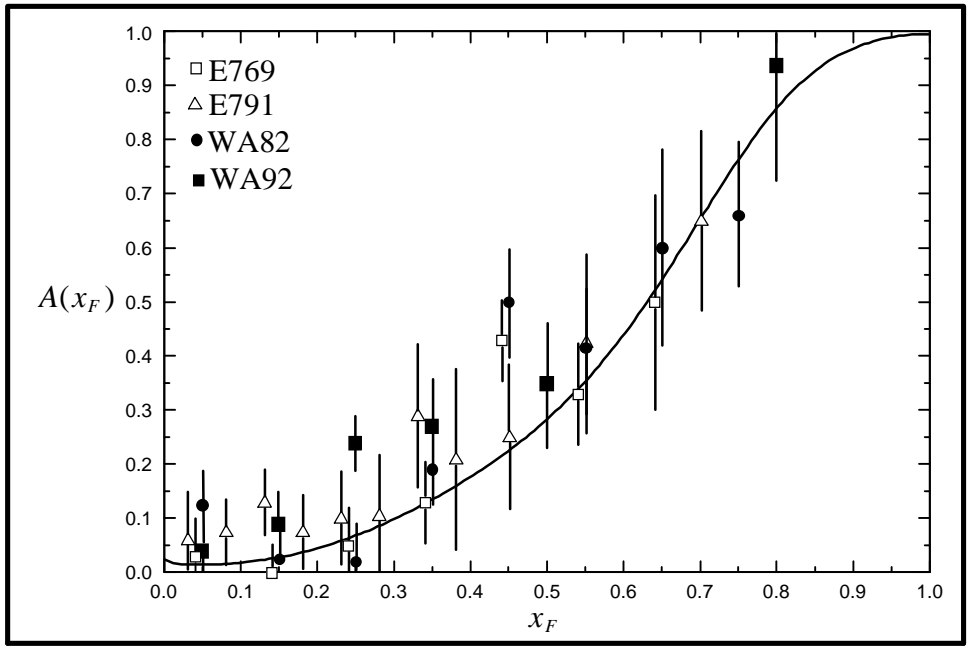
\item[{\bf Fig. 4}] Comparison of Eq. (19) for the asymmetry $A(x_F)$ with
experimental data  \cite{e769,ait,wa82,wa92}.

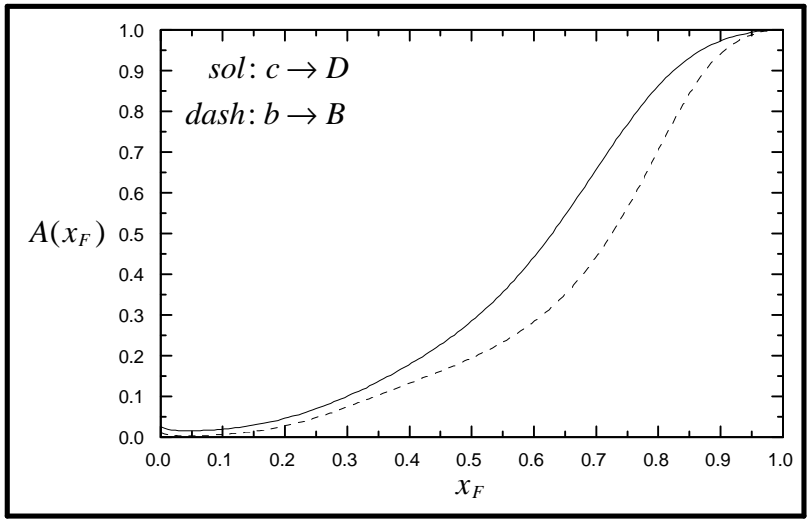
\item[{\bf Fig. 5}]  $D^{\mp}$ and $B^{\mp}$ asymmetry as a function of $x_F$.

\end{itemize}

\end{document}